\documentclass[A4paper,12pt]{article}
\usepackage[T1]{fontenc}
\usepackage[english]{babel}
\usepackage[cp1250]{inputenc}
\usepackage{epsfig}
\usepackage{amsfonts}
\usepackage{fancyhdr}
\usepackage{syntonly}
\usepackage{graphicx}

\begin{document}
\selectlanguage{english}

\begin{titlepage}
\begin{center}
\vspace*{3cm}

\begin{title}
\bold {\Huge Remarks on the particle multiplicities\\
\vspace{0.5cm}
 at LHC energies
 }
\end{title}

\vspace{2cm}

\begin{author}
\Large K. FIA{\L}KOWSKI\footnote{e-mail address:
fialkowski@th.if.uj.edu.pl}, R. WIT\footnote{e-mail address:
romuald.wit@uj.edu.pl}

\end{author}

\vspace{1cm}

{\sl M. Smoluchowski Institute of Physics\\ Jagellonian University \\

30-059 Krak{\'o}w, ul.Reymonta 4, Poland}

\vspace{2cm}

\begin{abstract}
Recent ALICE data for the multiplicity distributions in the central
rapidity bins at LHC energies are compared with the results from two
default versions of the PYTHIA 8 generator. We find that, contrary
to the earlier versions of PYTHIA, the model overestimates the
increase of average multiplicity with energy. Tuning two of the
model parameters one obtains reasonable agreement with data. The
dependence of the normalized moments of the distribution on the
rapidity bin width and on energy is also qualitatively correct.

\end{abstract}

\end{center}

\vspace{2cm}


\end{titlepage}

\section{Introduction}

The data on the charged particle density in the central rapidity
bins were recently published by the ALICE collaboration
\cite{ALICE},\cite{ALICE2} (similar data are available from the CMS
collaboration \cite{CMS}, whereas the ATLAS data are limited to
relatively high $p_T$).  The energy dependence of the average
multiplicity in the central unit of pseudorapidity found for the new
range seems to be stronger than in the lower energy data and the
observed increase is claimed to exceed significantly that expected
from the PYTHIA 6 and PHOJET event generators.
\par
In this note we show that the situation is very different for the
PYTHIA 8 generator (written in C++). The default versions of this
generator yield an increase of average multiplicity which is not
only much stronger than that from the PYTHIA 6, but also stronger
than the increase seen in the data. Tuning just two of the model
parameters one obtains a good description of data. Since the PYTHIA
8 generator is still in the development stage, we use not only the
most recent 8.135 version, but compare it with an earlier 8.107
version which describes differently the diffractive processes. These
versions differ in the predictions for the energy dependence of the
particle density in the inelastic events. We discuss shortly the
choices of event classes used in different analyses of the ALICE
data. We compare also the model results with the data for average
multiplicity and the normalized moments of the multiplicity
distributions in the rapidity bins of various width for two lower
LHC energies and find a reasonable agreement of the tuned version of
PYTHIA 8.135 with the data.
\par
    In the following section we give the details of our
generation procedure
    and of the definitions of quantities to be compared with data. Then we present the
    results and compare them with the ALICE data. Short conclusions
    are contained in the last section.

\section{Procedures and definitions}

In this note we are using the recent C++ versions of the PYTHIA
generator: 8.107 and 8.135  \cite{SMS}, \cite{SMS2}. Samples of 100 000 minimum bias events
are generated for the pp collisions at LHC
energies. To discuss the influence of the changes in the diffractive
component \cite{NAV} we generate separately the single and double
diffractive events, as well as the full samples of inelastic events.
\par
In the ALICE data at $900$ GeV and $2.36$ TeV \cite{ALICE} a table
of the average charged particle multiplicities is given for the CM
pseudorapidity interval $\mid \eta \mid <0.5$ (i.e., it is the
density in pseudorapidity). Three categories of events are
considered: inelastic, non-single-diffractive and inelastic with
$N_{ch}>0$. The results are compared with the old UA5 measurements
and the model calculations from three models: Quark Gluon String
Model, three versions of PYTHIA 6 (different "tunings") and PHOJET.
There is no significant discrepancy between two sets of data, and
the model results spread around the experimental values, although
the increase with energy is systematically underestimated.
\par
The data at $7$ TeV \cite{ALICE2} are compared with the lower energy
data and with the predictions of PYTHIA and PHOJET generators just
for one class of events: inelastic with $N_{ch}>0$. Here one defines
the density slightly differently: it is a half of the average
multiplicity in the pseudorapidity interval $\mid \eta \mid <1$. One
sees a clear discrepancy between all the models and data.
\par
Let us note that the different definitions of the sets of data to be
compared with models rely in a different degree on the models
themselves. To get the "non-single-diffractive" sample one has to
remove from the data the single diffractive events, which obviously
cannot be done in a model independent way. On the other hand, the
"inelastic" sample is measured in the model independent way, but the
model calculations treat the diffractive events in a different way
than the non-diffractive ones. Thus the (dis)agreement between the
models and data depends always on the description of both
diffractive and non-diffractive events.
\par
The single and double diffraction are described differently in the
"old" versions of PYTHIA (PYTHIA 6 and PYTHIA 8 up to the 8.130
version) and in PYTHIA 8.135.
At $7$ TeV CM energy the charged multiplicity distributions
for the single diffractive component from PYTHIA 6.414 is rather sharply cut around the
multiplicity of forty charged particles (in full phase space);
from PYTHIA 8.135 one gets a long tail extending beyond the
multiplicity of one hundred \cite{NAV}. We have checked that the
PYTHIA 8.107 version provides similar results as PYTHIA 6.414. Analogous
effect appears for double diffraction. Therefore the "INEL>0" class
contains different contributions of the diffractive events in the
two versions of PYTHIA, although the total cross sections for
non-diffractive, single diffractive and double diffractive
interactions are the same for both versions at all energies.

\section{PYTHIA 8 and the ALICE data}

The most recent ALICE paper \cite{ALICE2} presents a table of
charged particle pseudorapidity density measured at central
pseudorapidity ($\mid \eta \mid <1$) for inelastic collisions having
at least one charged particle in the same region. The experimental
data are compared with PHOJET and three versions of PYTHIA 6
generators at three LHC energies. All the generators underestimate
the density with the exception of PYTHIA with ATLAS-CSC tune at
$900$ GeV. Moreover, all generators underestimate significantly the
increase of density with energy, predicting the $15-18\%$ increase
between $900$ and $2360$ GeV, where the experimental increase is
$23.3(+1.1/-0.7)\%$, and the $33-48\%$ increase between $900$ GeV
and $7$ TeV, where the data show a $57.6(+3.6/-1.8)\%$ increase.
\par
We have performed the same calculation using the default versions of
PYTHIA 8.107 and 8.135 generators. The results are shown in  Table \ref{PYTH1}
and in Fig.1.

\begin{table}[h]
\caption{Central density: data and the results for two versions of PYTHIA}
\label{PYTH1}
\vspace{10pt}
\begin{center}
\begin{tabular}{||c|c|c|c|c||}
 \hline
 \hline
       Energy (TeV)& ALICE & PYTHIA 8.107 & PYTHIA 8.135&PYTHIA8.135 tuned\\
\hline \hline
          0.9 & 3.81(1)(7) & 3.81         &  4.00   & 3.83     \\
\hline
  2.36 & 4.70(1)(11) &4.93& 5.36 & 4.67 \\
\hline
     7.0 & 6.01(1)(20)  & 6.58& 7.66 & 5.95 \\
\hline
\hline

\end{tabular}
\end{center}
\end{table}

\begin{figure}[h]
\centerline{ \epsfig{figure=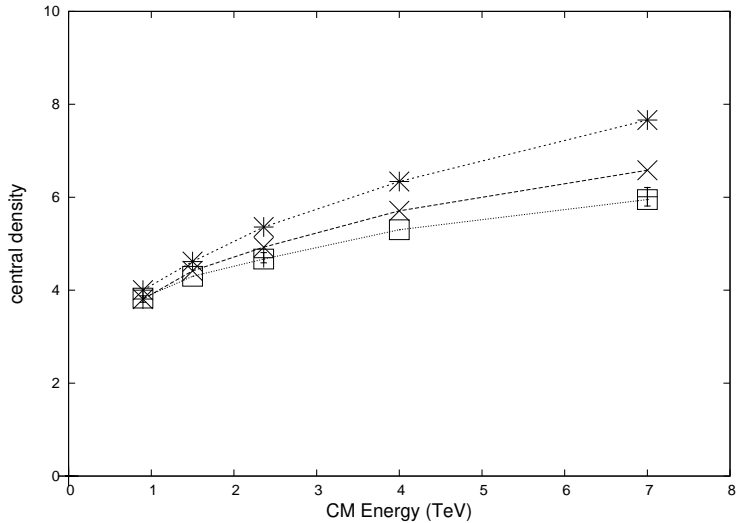,height=7.0cm}}
\caption{\footnotesize \label{Average} The central density in
pseudorapidity for the ALICE $pp$ data (crosses with error bars),
PYTHIA 8.107 (x signs), PYTHIA 8.135 default (asterisks) and PYTHIA
8.135 tuned (squares) as a function of the CM energy. The model
prediction points are connected by lines to guide the eye.}
\end{figure}

We see that the PYTHIA 8.107 results agree perfectly with data at
$900$ GeV and show faster increase with energy than the data: the
corresponding increase between $900$ and $2360$ GeV is $29.4\%$, and
between $900$ GeV and $7$ TeV it is $72.8\%$. The disagreement is
even stronger for the PYTHIA 8.135 generator: here all the values
are significantly higher than the data, and the increase is $33.9\%$
between $900$ and $2360$ GeV and $91.5\%$ between $900$ GeV and $7$
TeV. Thus instead of a significant underestimation, we get now a
significant overestimation of the experimental increase of the
central density with energy. However, by tuning just two of the
PYTHIA 8.135 parameters we are able to reproduce the experimental
values from ALICE. These two parameters $ecmRef$ and $ecmPow$ are
determining the low $p_T$ regularization of the (divergent) QCD
cross section by introduction of a factor
$$F(p_T)=\frac{p_T^4}{(p_{T0}^2+p_T^2)^2}$$
where
$$p_{T0}=pT0Ref\Big(\frac{ecmNow}{ecmRef}\Big)^{ecmPow}$$
and $ecmNow$ is the CM energy in GeV. For the $pT0Ref$ parameter we
take the default value of $2.0$, whereas $ecmRef$ is changed from
$1960.0$ to $1250.0$ and $ecmPow$ from $0.16$ to $0.26$. With these
values of parameters we reproduce within errors the experimental
values, as seen in the last column of Table 1 and in Fig.1.
\par
Obviously, more work is needed to tune PYTHIA 8 for the global
description of LHC data. However, we may safely say that we have
found a counterexample to the claim that the increase of density
with energy at LHC range is faster than expected in the commonly
used generators. Moreover, since PYTHIA 8 (as well as PYTHIA 6) is
based on the string fragmentation mechanism and not on the
thermodynamical picture, our results suggest that it is not
necessary to invoke the power-like thermodynamical increase of
multiplicity with energy to describe the LHC data. Tuning the above
mentioned two parameters one is able to reproduce this increase
quite well.
\begin{table}[!h]
\caption{Average multiplicities and moments for three choices of
rapidity bin from ALICE and two versions of PYTHIA 8.135 at $0.9$
TeV. The numbers in parentheses denote the statistical and
systematic errors.} \label{Moments} \vspace{10pt}
\begin{center}
\begin{tabular}{||c|c|c|c|c||}
\hline
 \hline
       Quantity& $ \eta $ range & ALICE & PYTHIA 8.135 default & PYTHIA 8.135 tuned \\
\hline
\hline
   $\overline n$&  $\mid \eta \mid <0.5$ &3.60(2)(11)&4.11&3.72  \\
\hline
   $\overline n$&  $\mid \eta \mid <1.0$ &7.38(3)(17)&8.35 & 7.58  \\
\hline
   $\overline n$&  $\mid \eta \mid <1.3$ &9.73(12)(19)&10.95 & 9.93 \\
\hline
  $c_2 $ & $\mid \eta \mid <0.5$ & 1.96(1)(6) & 1.81 & 1.81 \\
\hline
  $c_2 $ & $\mid \eta \mid <1.0$ & 1.77(1)(4) & 1.66 & 1.64 \\
\hline
  $c_2 $ & $\mid \eta \mid <1.3$ & 1.70(3)(7) & 1.62 & 1.60 \\
\hline
    $c_3$ & $\mid \eta \mid <0.5 $ &5.35(6)(31)&  4.70 & 4.75 \\
\hline
 $c_3$ & $\mid \eta \mid <1.0 $ &4.25(3)(20)&  3.92 & 3.89 \\
\hline
 $c_3$ & $\mid \eta \mid <1.3 $ &3.91(10)(15)&  3.71 & 3.67 \\
\hline
  $c_4$& $\mid \eta \mid <0.5$ & 18.3(4)(1.6)& 15.66 & 16.34 \\
\hline
  $c_4$& $\mid \eta \mid <1.0$ & 12.6(1)(9)& 11.76 & 11.87 \\
\hline
  $c_4$& $\mid \eta \mid <1.3$ & 10.9(4)(6)& 10.74 & 10.79 \\
 \hline
 \hline
\end{tabular}
\end{center}
\end{table}

\begin{table}[!h]
\caption{Data and models as in Table 2 at  $2.36$ TeV. }
\label{Moments2}
\vspace{10pt}
\begin{center}
\begin{tabular}{||c|c|c|c|c||}
\hline
 \hline
       Quantity& $ \eta $ range & ALICE & PYTHIA 8.135 default & PYTHIA 8.135 tuned \\
\hline
\hline
  $\overline n$ & $\mid \eta \mid <0.5$ &4.47(3)(10) & 5.54& 4.55 \\
\hline
  $\overline n$ & $\mid \eta \mid <1.0$ &9.08(6)(29) & 11.25 & 9.23 \\
 \hline
  $\overline n$ & $\mid \eta \mid <1.3$ &11.86(22)(45) & 14.75 & 12.09 \\
\hline
  $c_2$ & $\mid \eta \mid <0.5$ & 2.02(1)(4) & 1.93 & 1.86 \\
\hline
  $c_2$ & $\mid \eta \mid <1.0$ & 1.84(1)(6) & 1.81 & 1.72 \\
\hline
  $c_2$ & $\mid \eta \mid <1.3$ & 1.79(3)(7) & 1.77 & 1.68\\
  \hline
  $c_3$ &  $\mid \eta \mid <0.5$ & 5.76(9)(26)& 5.42 & 5.07 \\
\hline
  $c_3$ &  $\mid \eta \mid <1.0$ & 4.65(6)(30)& 4.72 & 4.27 \\
\hline
  $c_3$ &  $\mid \eta \mid <1.3$ & 4.35(16)(33) & 4.51 & 4.05 \\
  \hline
  $c_4$ & $\mid \eta \mid <0.5$ & 20.6(6)(1.4)& 19.2 & 17.84 \\
\hline
  $c_4$ & $\mid \eta \mid <1.0$ & 14.3(3)(1.4)& 15.3 & 13.53 \\
\hline
  $c_4$ & $\mid \eta \mid <1.3$ & 12.8(7)(1.5)& 14.2 & 12.38 \\
\hline
\hline
\end{tabular}
\end{center}
\end{table}

\par
In the first paper \cite{ALICE} the ALICE Collaboration has shown
the data for "non-single diffractive" events for two energies ($0.9$
and $2.360$ TeV) and for three choices of the central rapidity bin
width. The average multiplicities, and the three lowest normalized
moments $c_i=<n^i>/<n>^i$ of the multiplicity distribution were
measured. These results were shown to agree with the old UA5 data at
$900$ GeV; no comparison with any model was presented there.
\par
In Table \ref{Moments} and Table \ref{Moments2} we present the
comparison of these results at $0.9$ TeV and $2.36$ TeV,
respectively, with those from the default and tuned versions of
PYTHIA 8.135.
\par
 For the default version of PYTHIA 8.135 the average multiplicity is always significantly too high.
 The tuned version is in perfect agreement with the data. The values of the moments are not too well
 reproduced, but only in one case the deviation from data is more than thrice the
 errors. The tuning in some cases increases this deviation, and in
 some cases reduces it. Let us stress here that the parameters were
 chosen to fit the data from Table 1, and not the data from Table 2 and Table 3.
 In any case, the qualitative features of the data, as the decrease of
 moments with the width of the rapidity bin and the increase with energy, are reproduced quite well
 in both versions of PYTHIA. Dedicated tuning of the parameters should improve the agreement with data.
\section{Conclusions}
We have investigated the ALICE data for the multiplicity
distributions in the central rapidity region using two default
versions of the PYTHIA 8 generator. We find that, in comparison with
data, the model overestimates the increase of central density with
energy, contrary to the older versions of the PYTHIA generator. The
increase is strongest for the 8.135 version, which includes the hard
diffractive processes. This shows that the string models are not
bound to underestimate the increase of multiplicity with energy seen
at LHC (as often suggested). Tuning the model parameters one obtains
a good agreement with data.
\par
The data for higher moments of the multiplicity distribution for two
lower LHC energies show similar features. None of the versions of
the generator reproduces the data really well, but the results fall
around the data. It is quite likely that a systematically tuned
version may describe the experimental results in a satisfactory way.

\section{Acknowledgments} We thank Torbjoern Sjoestrand for suggesting the choice of parameters to be tuned. We are
grateful to Andrzej Kotański for  helpful remarks.

\end{document}